# Geometric Scaling of the Current-Phase Relation of Niobium Nano-Bridge Junctions

Yue Wang[1,2], Lei Chen[1,2,*], Yinping Pan[1], Denghui Zhang[1,2], Shujie Yu[1,2], Guangting Wu[1,2], Xiaoyu Liu[1], Ling Wu[1], Weifeng Shi[1], Guofeng Zhang[1], Lu Zhang[1,2], Wei Peng[1,2], Jie Ren[1,2], Zhen Wang[1,2,3,*]

[1] National Key Laboratory of Materials for Integrated Circuits, Shanghai Institute of Microsystem and Information Technology (SIMIT), Chinese Academy of Sciences, Shanghai 200050, China

[2] University of the Chinese Academy of Sciences, Beijing 100049, China

[3] School of Physical Science and Technology, Shanghai Tech University, Shanghai 200031, China

*Corresponding author: leichen@mail.sim.ac.cn, zwang@mail.sim.ac.cn

**Abstract:** The nano-bridge junction (NBJ) is a type of Josephson junction that is advantageous for the miniaturization of superconducting circuits. However, the current-phase relation (CPR) of the NBJ usually deviates from a sinusoidal function which has been explained by a simplified model with correlation only to its effective length. Here, we investigated both measured and calculated CPRs of niobium NBJs of a cuboidal shape with a three-dimensional bank structure. From a sine-wave to a saw-tooth-like form, we showed that deviated CPRs of NBJs can be described quantitatively by its skewness $\Delta\theta$. Furthermore, the measured dependency of $\Delta\theta$ on the critical current $I_0$ from 108 NBJs turned out to be consistent with the calculated ones derived from the change in geometric dimensions. It suggested that the CPRs of NBJs can be tuned by their geometric dimensions. In addition, the calculated scaling behavior of $\Delta\theta$ versus $I_0$ in three-dimensional space was provided for the future design of superconducting circuits of a high integration level by using niobium NBJs.

**Keywords:** nano-bridge junction, Josephson effect, current-phase relation, geometric scaling, superconducting circuit

The superconducting circuits have attracted rising attention for their ultra-high speed and ultra-low energy consumption.[1-5] However, the integration level of superconducting circuits remains far behind the complementary metal-oxide-semiconductor (CMOS) technology.[6, 7] The nano-bridge junction (NBJ) shaped as a superconducting constriction is one type of Josephson junction (JJ) that is suitable for miniaturization integration by its very nature.[8-10] The nanoscale superconducting quantum interference device (SQUID) comprising two parallel NBJs has made important contributions to the study of nano-magnetism[11-16] and led to the development of scanning SQUID microscopes.[17, 18] Several types of miniaturized superconducting memory cell based on NBJs have been developed for superconducting computers recently.[19-21] Furthermore, qubits made of superconducting NBJs have been demonstrated to increase the integration level of quantum circuits.[22] The performances of all these devices are mainly dominated by the shape of the current-phase relation (CPR) of NBJs, known as the DC Josephson equation. Unlike the CPR of the superconducting tunneling junction, the CPR of the NBJ usually deviates from a sinusoidal shape.[10, 23, 24]

Kulik and Omelyanchuk modeled a small bridge as a one-dimensional filament with a length much smaller than the superconducting coherence length $\xi_0$ and obtained the deviation of CPR from a sinusoidal form as a function of temperature.[25] For bridges of intermediate size, Likharev et al. correlated the deviation of the CPR to the ratio between the effective length $l_{eff}$ and $\xi_0$, where $l_{eff}$ was intuitively approximated as the greater of the length and width.[10] Experimentally, the CPR deviation of a small bridge induced by temperature has been measured for both an aluminum point contact and a short niobium (Nb) bridge.[26, 27] The CPRs of several non-traditional JJs have also been measured to reveal their intrinsic physical properties.[28-30] However, most experimental studies on CPRs only considered the case that the junction is much smaller than $\xi_0$, which is a challenging regime for NBJs by using common clean-room technology. The deviated CPRs of NBJs with the geometric dimensions comparable to or even larger than the $\xi_0$ have not yet been carefully studied.

In the present study, we investigated the geometric scaling behavior of the CPR of the Nb NBJ with a three-dimensional (3D) bank structure.[31, 32] The phase gradient reduced appreciably in the thick banks, which allowed us to study the Josephson effect purely induced by the geometric dimensions of the bridge deck. We showed that the deviated shape of CPRs of NBJs can be schematically quantified by the skewness $\Delta\theta$. Furthermore, the observed dependency of $\Delta\theta$ on critical current $I_0$ of NBJs is consistent with the calculated ones corresponding to the change of geometric dimensions. In addition, we presented the scaling pattern of $\Delta\theta$ vs $I_0$ with the geometric size of NBJs in three-dimensional space. It provides a future reference in designing superconducting circuits by using Nb NBJs.

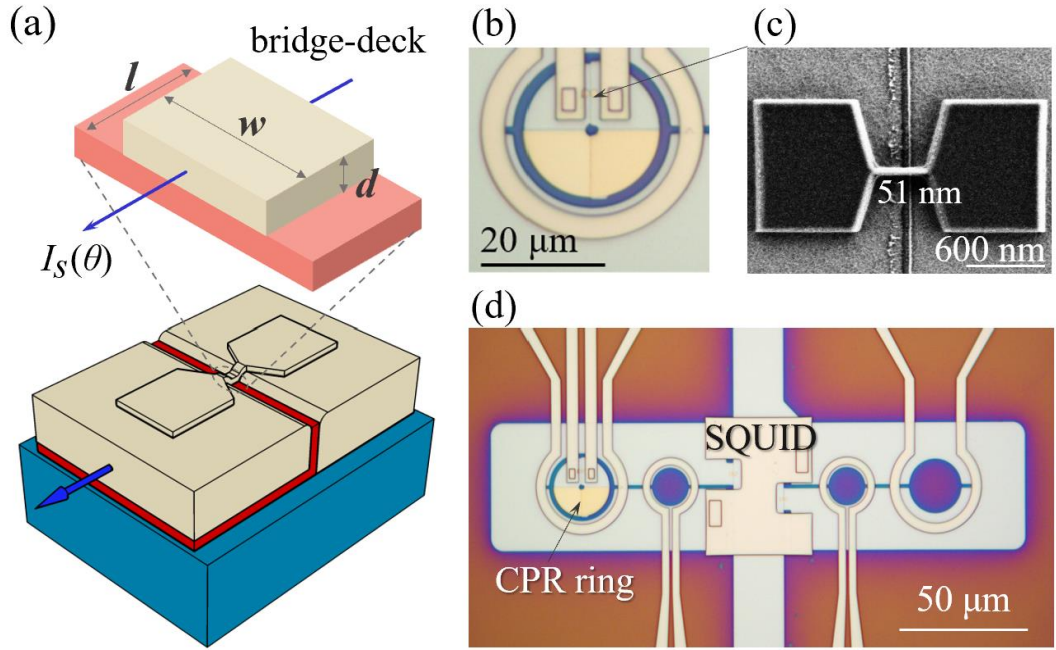

Figure 1. (a) Schematic drawing of the 3D nano-bridge junction. The deck of the bridge was modeled as a cuiboid with width $w$, thickness $d$, and length $l$. (b) Optical micrograph of the CPR ring. (c) Scanning electron micrograph of the 3D nano-bridge. (d) Optical micrograph of the CPR measurement circuit. The nano-bridge was connected to the CPR ring, whose enclosed flux was measured using a gradiometric SQUID.

**Results**

The 3D structure of the Nb NBJ is sketched in Figure 1a. The cuboidal bridge deck links two thick banks separated by an insulated slit. The thickness of each bank is about 150 nm, and the breadth of the insulated slit is approximately 20 nm. The deck length $l$ is then determined by the breadth of the insulated slit. The deck width $w$ is defined by the e-beam lithography. The deck thickness $d$ is in principle determined by the film thickness of the bridge layer, which is 8 nm. The scanning electron micrograph in Figure 1c shows the top view of the nano-bridge. The 3D Nb nano-bridge structure was fabricated following the procedure in the literature.[32] Details of the fabrication are described in the method section. Figure 1c shows that a rough notch at the top of the insulated slit is left by the lift-off step in the fabrication process. Although the deposition thickness of the bridge layer is 8 nm, the actual $d$ is random distributed over the silicon wafer because the deposition uniformity on the wafer is degraded by the roughness over the notch (see Supporting Information Figure S2).

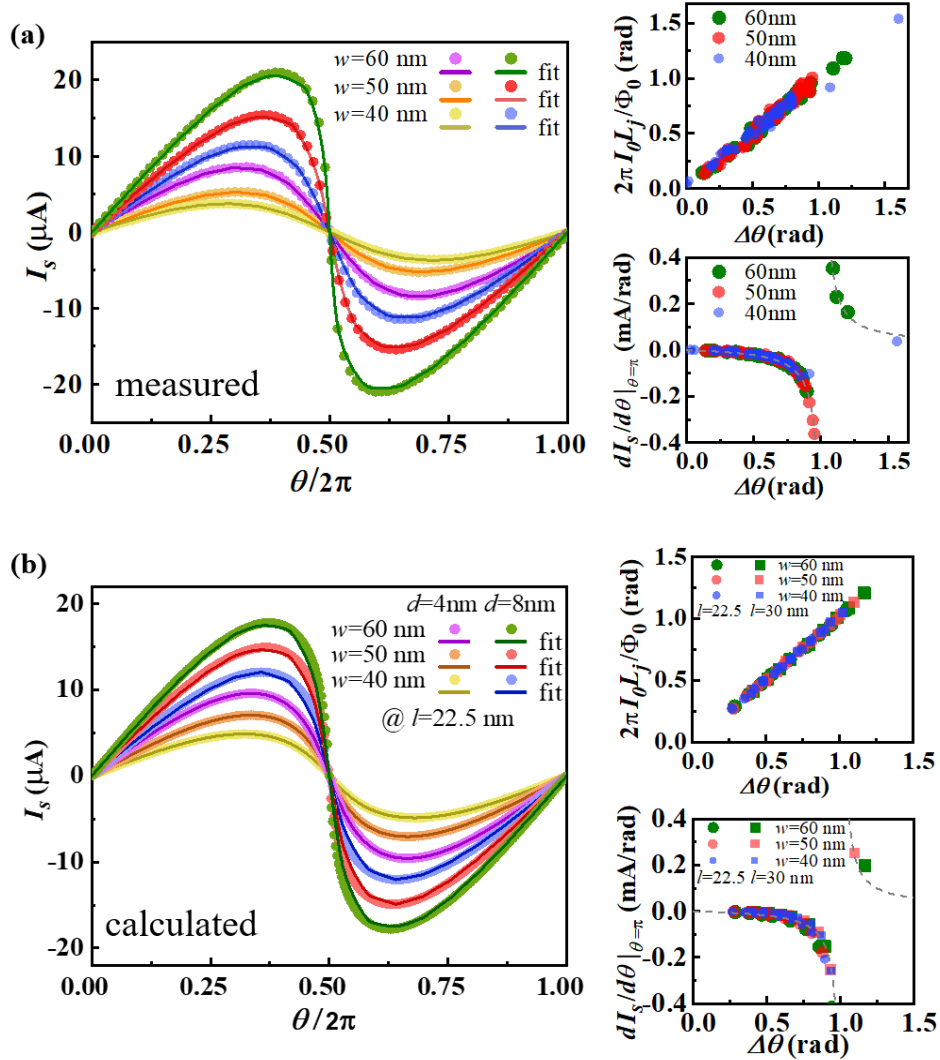

Figure 2. CPRs measured (a) and calculated (b) for several samples with $w$ = 40, 50, and 60 nm. The lines were fitted using equation (1) with fitting parameters $I_0$ and $L_j$. The upper-right panels in (a) and (b) show the relation between the skewness $\Delta\theta$ and the fitting parameter $I_0 L_j$ obtained from measured and calculated CPRs, respectively. Both plots satisfy $\Delta\theta = 2\pi L_j I_0/\Phi_0$. The lower-right panels in (a) and (b) show the relation between the derivative of the CPR at $\theta = \pi$ and $\Delta\theta$ from the measured and calculated CPRs, respectively. Both plots satisfy $dI_s/d\theta|_{\theta=\pi} = I_0/(\Delta\theta - 1)$.

The NBJ was connected to a superconducting ring (Figure 1b) to measure the CPR as shown in Figure 1d. The CPR measurements were performed in both flux-bias[29] and current-bias[30] methods to avoid mistakes during the data analysis. Both methods obtained identical CPR curves and thus confirmed each other. For comparison, we also calculated the CPRs of the nano-bridges in the 3D structure by solving the Ginzburg–Landau equations in the finite element method.[33] The bridge was modeled as a cuboidal shape of length $l$, width $w$, and thickness $d$. Typical measured and calculated CPR curves are shown in Figure 2 a and b, respectively. Details on the CPR measurement and calculation can be found in the method section.

In consideration of uncertainty in fabrication, we measured the CPR for 108 samples fabricated on the same silicon wafer (i.e., 36 samples for each of $w$ = 40, 50, and 60 nm, see Supporting Information Figure S1). The measurements were made in liquid helium at a temperature of 4.2 K. Because of the nonuniformity of the bridge deck over the wafer, the measured CPR changed from a sinusoidal to a saw-tooth shape. Figure 2a shows several typically measured CPR curves of each width. For comparison, the calculated CPRs of the NBJs with various geometric dimensions, obtained by numerically solving the Ginzburg–Landau equations, are plotted in Figure 2b. More data can be found in Supporting Information Figure S3. We assumed that the superconducting NBJs can be modeled as a JJ connected in series with an inductor $L_j$. Then, the total phase $\theta$ drops partially across an inductor with $\theta_L = 2\pi L_j I_s/\Phi_0$, and partially across a JJ with sinusoidal CPR, $I_0\sin(\theta-\theta_L)$, where $I_0$ is the critical current of the JJ. Thus, we can use the implicit function $I_s(\theta) = I_0\sin[\theta - 2\pi L_j I_s(\theta)/\Phi_0]$ to fit both measured and calculated CPRs.

In order to quantify the change in the CPR, we define the skewness as $\Delta\theta = \theta_{max} - \pi/2$, where $\theta_{max}$ corresponds to $I_s(\theta_{max}) = I_0$. $I_0 L_j$ and the CPR derivative $dI_s/d\theta$ at $\theta = \pi$ as a function of $\Delta\theta$ are plotted on the right of Figure 2a and 2b for the measured and calculated CPRs, respectively. Consequently, the as-defined skewness $\Delta\theta$ of both the measured and calculated CPR curves satisfies the equations $\Delta\theta = 2\pi I_0 L_j/\Phi_0$ and $dI_s/d\theta|_{\theta=\pi} = I_0/(\Delta\theta - 1)$. Therefore, the skewness $\Delta\theta$ as a normalized inductance can be used to quantify the Josephson effect of NBJs. The normalized CPR of NBJs becomes:

$$i_s(\theta) = \sin[\theta - i_s(\theta)\Delta\theta] \qquad (1)$$

where $i_s(\theta) = I_s(\theta)/I_0$. A reduction in $\Delta\theta$ suggests an enhancement of the Josephson effect of NBJs. When $\Delta\theta < 1$ rad, the derivative of the CPR at $\theta = \pi$ is negative, the CPR is single-valued, and the nano-bridge junction acts more like a JJ. When $\Delta\theta > 1$ rad, the derivative of the CPR at $\theta = \pi$ becomes positive, the CPR becomes multi-valued, and the nano-bridge junction acts more like a linear inductor.

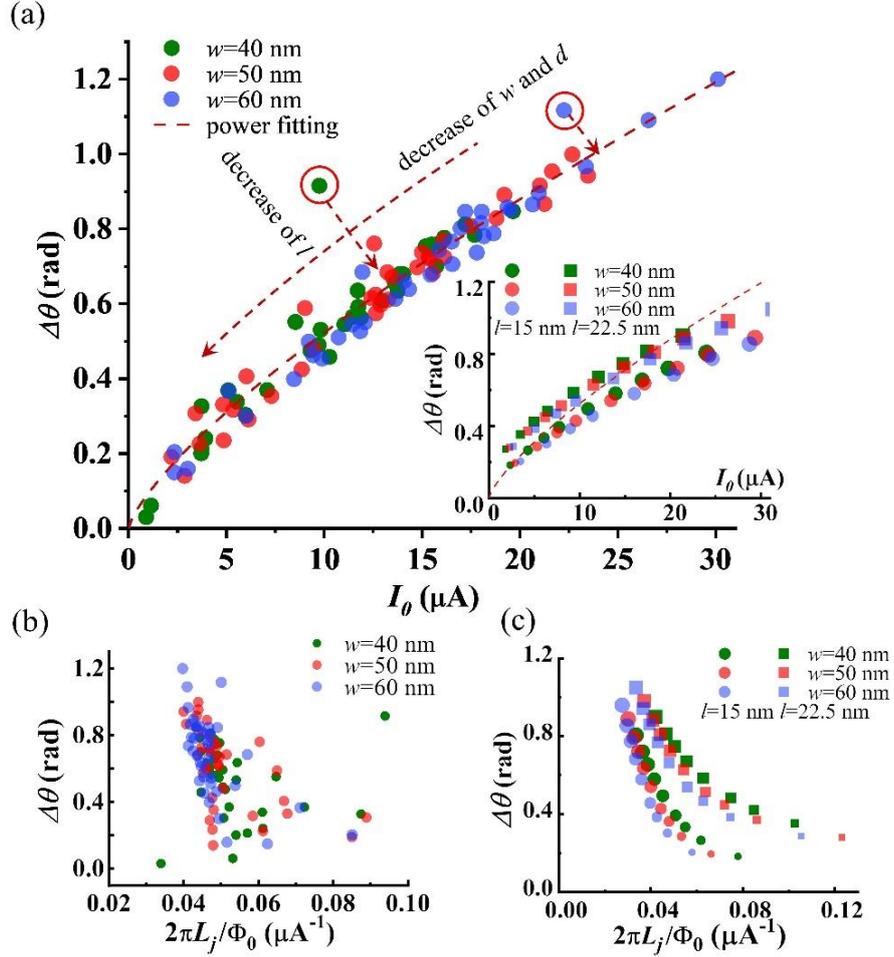

Figure 3. (a) Skewness $\Delta\theta$ as a function of $I_0$ obtained from the measured CPR for 36 samples plotted on one graph. Dark green, light red, and light blue show results for nano-bridges with $w$ = 40, 50, and 60 nm, respectively. The dash line is the fitting curve by a power function $\Delta\theta \propto I_0^{3/4}$. The inset shows $\Delta\theta$ as a function of $I_0$ from the calculated CPR. Circles and squares are results for samples with $l$ = 15 and 22.5 nm respectively. The increase in $d$ from 1 to 15 nm is represented by the symbol size. $\Delta\theta$ as a function of $L_j$ obtained from (b) measured and (c) calculated CPRs. The change in $d$ from 1 to 15 nm in (c) is represented by the symbol size.

Given the evidence that the shape of the CPR of the NBJ depends on $\Delta\theta$, the combination of $\Delta\theta$ and $I_0$ is important for application in different types of superconducting circuit. The skewness $\Delta\theta$ as a function of $I_0$ obtained from the measured CPR for all 108 samples is plotted in Figure 3a. Dark green, light red, and light blue show results for the nano-bridges with $w$ = 40, 50, and 60 nm, respectively. Most of the data fall on a single dashed curve of the power function $\Delta\theta \propto I_0^{3/4}$. For comparison, we show $\Delta\theta$ as a function of $I_0$ obtained from calculated CPRs in the inset of Figure 3a. Dark green, light red, and light blue again show the results for nano-bridges with $w$ = 40, 50, and 60 nm, respectively. In the calculation model, $w$, $d,$ and $l$ of the nano-bridge are all defined. Circles and squares show results for $l$ = 15 and 22.5 nm respectively. The increase in $d$ from 1 to 15 nm is represented by a change in symbol size. The dependency curves of $\Delta\theta$ on $I_0$ scaled with $w$ and $d$ overlap for the same $l$. Then the increase in $l$ from 15 to 22.5 nm shifts the whole dependency curve upward slightly. The calculations and measurements show comparable dependency in Figure 3a. This suggests that both $I_0$ and $\Delta\theta$ decrease monotonically with the scaling down of $w$ and $d$ in a consistent way. Therefore, the CPRs transform along the dependency curve when the NBJ expands from a quasi-one-dimensional (-1D) junction into a 3D geometry with finite $w$ and $d$, as shown in Figure 3a.

The scaling of CPRs with $w$ and $d$ turned out to be on similar dependency curves. Then the increase in $l$ shifts the whole curve upward slightly. It is also confirmed that the as-fabricated $l$ of the most junctions is approximately 20 nm. There are two points (indicated by red circles) far from the dashed line in Figure 3a. The as-fabricated $l$ of those two NBJs is unintentionally longer than that of the other NBJs. This difference is likely due to occasional poor electrical contact between the bridge layer and bank layer. The as-fabricated $d$ of NBJs is distributed randomly for all three widths. Additionally, $\Delta\theta$ as a function of $L_j$ obtained from measured and calculated CPRs is plotted in Figure 3b and 3c, respectively. $\Delta\theta$ and $L_j$ are negatively correlated with the variation of $w$ and $d$ for both measured and calculated CPRs. It is worth noting that the measured $\Delta\theta$ is slightly greater than the calculated ones for junctions with higher $I_0$. It is because that the magnetic field generated by the nano-bridge junctions is neglected in the calculation in order to avoid time dependent terms. The self-inductance was not included in the calculated CPR. It will cause a small discrepancy between measured and calculated $\Delta\theta$ at the higher $I_0$. However, the discrepancy will not change the shape of the dependency curves. Both measured and calculated CPRs showed similar dependencies of $\Delta\theta$ as a function of $I_0$ and $L_j$, which suggests that the measured and calculated results are consistent with each other.

The average $I_0$ and $\Delta\theta$ are obtained from the CPRs of the same $w$ because $I_0$ and $\Delta\theta$ are randomly distributed for the 36 NBJs. Figure 4a and 4b present the average $I_0$ and $\Delta\theta$ at $w = 40$, 50, and 60 nm as blue squares. The average $I_0$ and $\Delta\theta$ obtained from the measured CPR clearly increase with $w$ and are well fit by the calculations presented as light-red circles. Both $I_0$ and $\Delta\theta$ increase with the increase of $w$ in a way that is similar to the scaling behavior of $d$. However, the behavior induced by the increase of $l$ is on the opposite. Unfortunately, $l$ of NBJs in the 3D bank structure that is determined by the breath of the insulating slit cannot be effectively modified in the fabrication process. The data deviating from the power-fitting curves in Figure 3a suggest that the calculated CPRs provide a good estimation of $\Delta\theta$ and $I_0$ with the change of $l$.

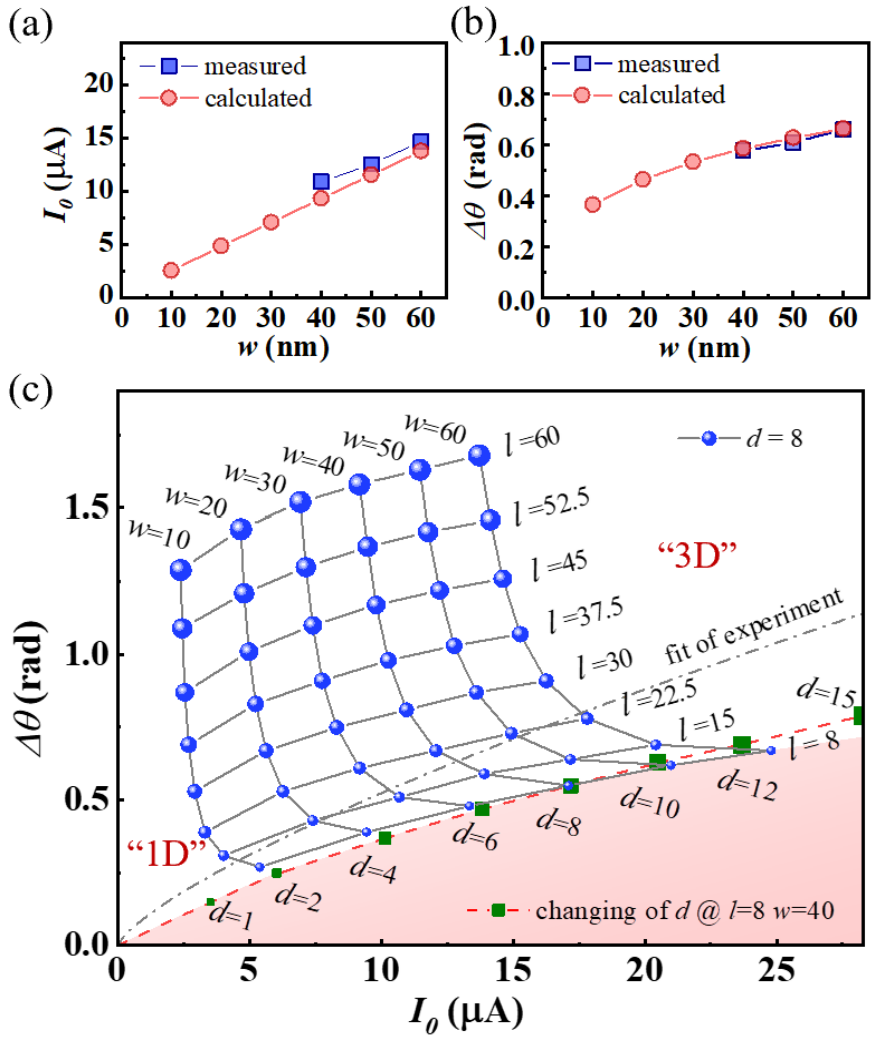

Figure 4. (a) Average measured (blue squares) and calculated (light-red circles) critical current $I_0$ as a function of $w$ of the NBJs. (b) Average measured (blue squares) and calculated (light red circles) skewness $\Delta\theta$ as a function of $w$ of the NBJs. (c) Lines connected by blue dots showing curves of $\Delta\theta$ as a function of $I_0$ with the change of $w$ and $l$. The almost vertical lines for constant $w$ (changing of $l$) and almost horizontal lines for constant $l$ (changing of $w$) form a mesh-like pattern. The dashed line connecting green squares is the curve obtained with the change of $d$. The dash-dot line is the fitting curve of $\Delta\theta \propto I_0^{3/4}$ obtained from the measured CPRs in Figure 3a. The light-red region below the bottom edge of the mesh indicates where the combination $\Delta\theta$ and $I_0$ is not attainable using a Nb NBJ with finite geometric dimensions. The units for $w$, $l$, and $d$ are all nanometers.

**Discussion**

We calculated the CPR for NBJs with $l$ ranging from 8 to 60 nm and $w$ ranging from 10 to 60 nm at $d = 8$ nm. In Figure 4c, the dependency of $\Delta\theta$ on $I_0$ for all calculated CPRs is presented by blue symbols. The symbol size increases with $l$. In contrast with the scaling behavior of $w$ and $d$, $I_0$ increases slightly with the scaling down of $l$. This is similar to the behavior of $L_j$ scaling with $w$ and $d$ as in Figure 3b and 3c. Therefore, curves of $\Delta\theta$ as a function of $I_0$ by changing $w$ and $l$ cross each other and form a mesh-like pattern. On the left edge of the mesh, $d = 8$ nm and $w = 10$ nm are both less than $\xi_0 = 15$ nm, and the NBJ can be approximated as a quasi-1D junction. $\Delta\theta$ increases with the increase of $l$. When $l > 3\xi_0$, $\Delta\theta > 1$ rad and the CPR becomes multi-valued, which agrees with the known CPR dependency on $l_{\text{eff}}$ from the 1D junction model.[10, 34, 35] For the 1D junction model, $\Delta\theta$ goes to zero with the decrease of $l_{\text{eff}}$, and the junction becomes a JJ with sinusoidal CPR. However, the mesh pattern starts to saturate at $l < \xi_0$. On the right edge of the mesh, when $w = 60$ nm, the combinations $\Delta\theta$ and $I_0$ represent the CPRs of NBJ in the 3D space. The decrease in $\Delta\theta$ saturates near the bottom edge of the mesh-like pattern with the decreasing of $l$ owing to the finite dimensions of $w$ and $d$. At the bottom edge of the mesh, the curve obtained with the increase of $w$ at $l = 8$ nm and $d = 8$ nm almost overlaps that obtained with the increase of $d$ at $l = 8$ nm and $w = 40$ nm, which is plotted as green squares connected by a dashed line. Therefore, $\Delta\theta$ can be further reduced by the decreasing of $w$ and $d$ along the bottom edge, but $I_0$ also drops accordingly. The decrease of $d$ by 2 nm and decrease of $w$ by 10 nm induced almost equal reduction in $\Delta\theta$. It suggests that the decrease of $d$ is more effective than decrease $w$ in improving the Josephson effect in NBJs. The light-red region under the mesh-like pattern is the region in which the combination of $I_0$ and $\Delta\theta$ is forbidden for a NBJ with finite geometric dimensions, because $l = 8$ nm is likely near the limit reachable using modern fabrication techniques.

In general, the Figure 4c suggests that the CPR of NBJs with finite $w$ and $d$ can be tuned close to a sinusoidal function with a small $\Delta\theta$ by decreasing $l$ down to a value less than $\xi_0/2$. Further decrease of $l$ will be very challenging in the fabrication and not help much to reduce $\Delta\theta$. Along the bottom edge of the mesh-like pattern in Figure 4c, $\Delta\theta$ can be further reduced by decreasing $w$ and $d$ with a quick drop in $I_0$. Although decrease of $w$ and $d$ plays a similar role in the reduction of $\Delta\theta$, decrease of $d$ is more effective and less challenging in the fabrication. However, the requirement of $I_0$ in real applications will prevent $\Delta\theta$ from zero. Therefore, the mesh-like pattern in Figure 4c provides a practical reference for estimating CPRs in the future design of superconducting devices and circuits by using NBJs.

**Conclusions:**

In summary, we measured and calculated the CPRs of NBJs in a 3D structure. Both measured and calculated CPRs indicate that the NBJ can be modeled as a JJ with a sinusoidal CPR connected in series with an equivalent inductor $L_j$. From a sine-wave to a saw-tooth like shape, the change in the CPR can be quantified by $\Delta\theta = 2\pi I_0 L_j/\Phi_0$, namely the skewness. Both $I_0$ and $L_j$ are further interconnected by the geometric scaling of the NBJs. By decreasing $w$ and $d$, the skewness $\Delta\theta$ is reduced with a quick drop in $I_0$ and a slow rise in $L_j$. The measured and calculated scaling behaviors of CPRs are consistent with each other. The deviation of CPRs from a sinusoidal function is restrained along the dependency curves when the model of NBJs expands from a 1D junction into a 3D geometry. The dependency of $\Delta\theta$ on $I_0$ with the changing of $w$ and $l$ forms a mesh-like pattern that bends at the bottom edge. Unlike the result for the 1D junction model, the decrease in $\Delta\theta$ saturates near the bottom edge with the decrease of $l$ owing to the finite dimensions of $w$ and $d$. The combination of $\Delta\theta$ and $I_0$ in the region below the bottom edge is not attainable using a Nb NBJ in reality. The mesh-like pattern can serve as a practical reference for estimating the CPRs required for designing different devices and circuits with a high integration level by using Nb NBJs in the future.

**Methods**

Sample preparation

Device fabrication was using a process adjusted from the 3D-NBJ fabrication process.[32] It comprises the following 10 main steps as sketched in Figure 5. (1) Deposition of a 150-nm first Nb layer on a silicon (Si) wafer with a 300-nm $SiO_2$ coating via DC magnetron sputtering. (2) Photolithography and reactive-ion etching (RIE) to pattern the first Nb film. The photoresist was kept on the sample after RIE. (3) Deposition of 30-nm $SiO_2$ by ion beam deposition (IBD) followed by deposition of a 180-nm second Nb layer by DC magnetron sputtering. (4) Patterning again the second Nb film through photolithography and RIE. (5) Lifting the photoresist, along with the film structure on top, by soaking the entire chip in acetone solution. A $SiO_2$ vertical insulating slit with a breadth of 20 nm forms between the first and second Nb structures. (6) Deposition of a 400-nm $SiO_2$ layer through plasma-enhanced chemical vapor deposition at 80 °C. (7) Planarization of the wafer by chemical mechanical polishing to obtain a relatively smooth cross-section. (8) Deposition of a thin third Nb layer. The Nb nano-bridges were then patterned across the $SiO_2$ slit through electron-beam lithography and RIE. (9) Deposition of a 200-nm insulating layer of $SiO_2$ by plasma-enhanced chemical vapor deposition at 80 °C. Conducting vias between the bottom base and the wiring Nb layer were then made through photolithography and RIE. (10) Deposition of a fourth Nb layer (wiring layer) through DC-magnetron sputtering. The current bias lines, input coils, SQUID feed-back coils, and contact pads were then patterned through photolithography and RIE.

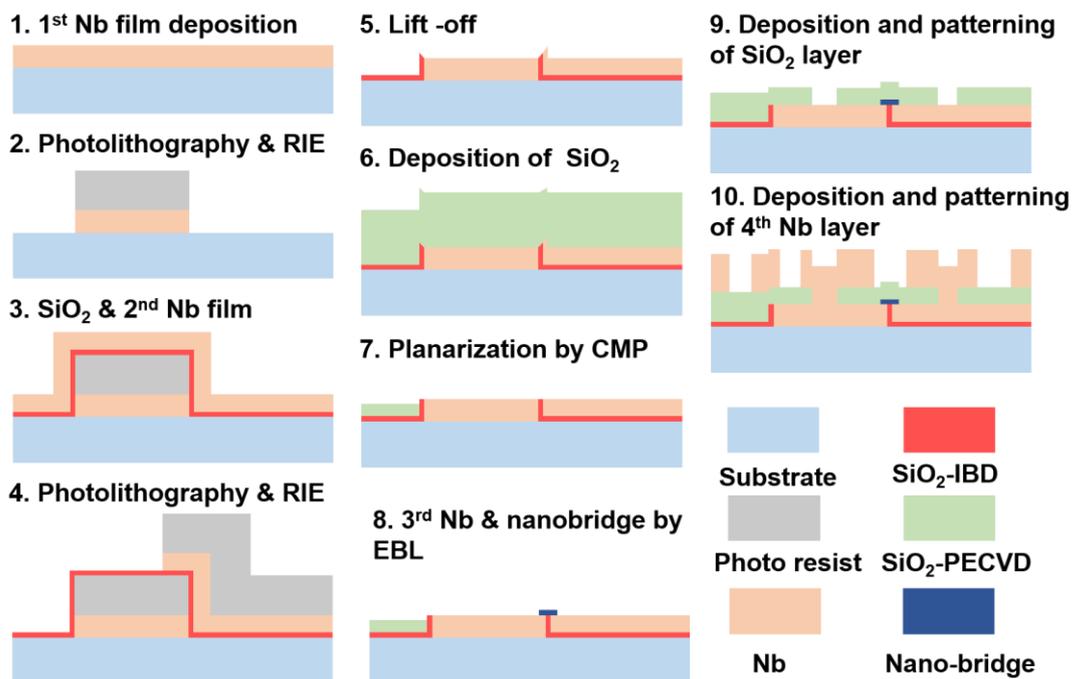

Figure 5. Main steps in fabricating devices for the CPR measurement.

Measurement of the CPR

Figure 6a is a schematic diagram for the CPR measurement setup. The input coil is counter-wounded to cancel the magnetic flux applied to the gradiometric SQUID.[36] The SQUID is flux locked by the feed-back coil to give a linear flux-to-voltage conversion $\Phi_f = \frac{V_{out}}{R_f} M_f$. The mutual inductance $M_f$ between the feed-back coil and the SQUID is characterized as $M_f \sim 7.143$ $\Phi_0$/mA. $R_f = 10$ kΩ is the resistance on the feedback line. The NBJ is connected into the CPR ring for measurements. We employ two methods to measure the CPR, namely the flux bias method and the current bias method. Typical measured curves by two methods are shown in Figure 6c and d, respectively.

In the flux bias method, we apply a current $I_{in}$ to the input coil to introduce a phase drop $\theta = 2\pi\Phi_a/\Phi_0 = 2\pi I_{in}/I_{p1}$ across the NBJ. $\Phi_a$ is the magnetic flux applied to the CPR ring, and $I_{p1} = 1.15$ mA is the period of $I_{in}$ corresponding to a $\Phi_0$, as shown in Figure 6c. The SQUID detects the magnetic flux generated by the supercurrent $I_s$ and the input coil. Therefore, $I_s = (\Phi_f - I_{in} M_{in\text{-}SQUID})/M_{s\text{-}SQUID}$. Here, $M_{in\text{-}SQUID} \sim 0.0463$ $\Phi_0$/mA is the mutual inductance between the input coil and the SQUID, which is deduced from the slope of the linear component in $\Phi_f$ as a function $I_{in}$. $M_{s\text{-}SQUID} \sim 1.732$ $\Phi_0$/mA is the mutual inductance between the CPR ring and the SQUID, which is characterized by applying the current $I_b$ to the CPR ring and measuring the flux response of the SQUID.

In the current bias method, a phase drop $\theta = 2\pi\Phi_s/\Phi_0 = 2\pi I_b/I_{p2}$ across the junction is introduced by injecting bias current $I_b$ to the CPR ring directly. As shown in Figure 6d, $I_{p2} = 0.456$ mA is the period of $I_b$ corresponding to a $\Phi_0$. The supercurrent $I_s$ flowing inside the CPR ring is measured as a total flux $\Phi_s$ coupled to the SQUID. Then $I_s = I_b - k\Phi_s/M_{s\text{-}SQUID}$, where $k = \frac{\Phi_f}{\Phi_s} = \frac{\Phi_p}{\Phi_0}$ is the flux coupling coefficient of the sample ring and SQUID loop, and $M_{s\text{-}SQUID} \sim 1.732$ $\Phi_0$/mA is the mutual inductance between the CPR ring and the SQUID, which is deduced from the linear slope of $\Phi_f$ as a function of $I_b$. As shown in Figure 6b, the deduced CPR curves from the flux-bias method and the current-bias method are identical and confirm each other.

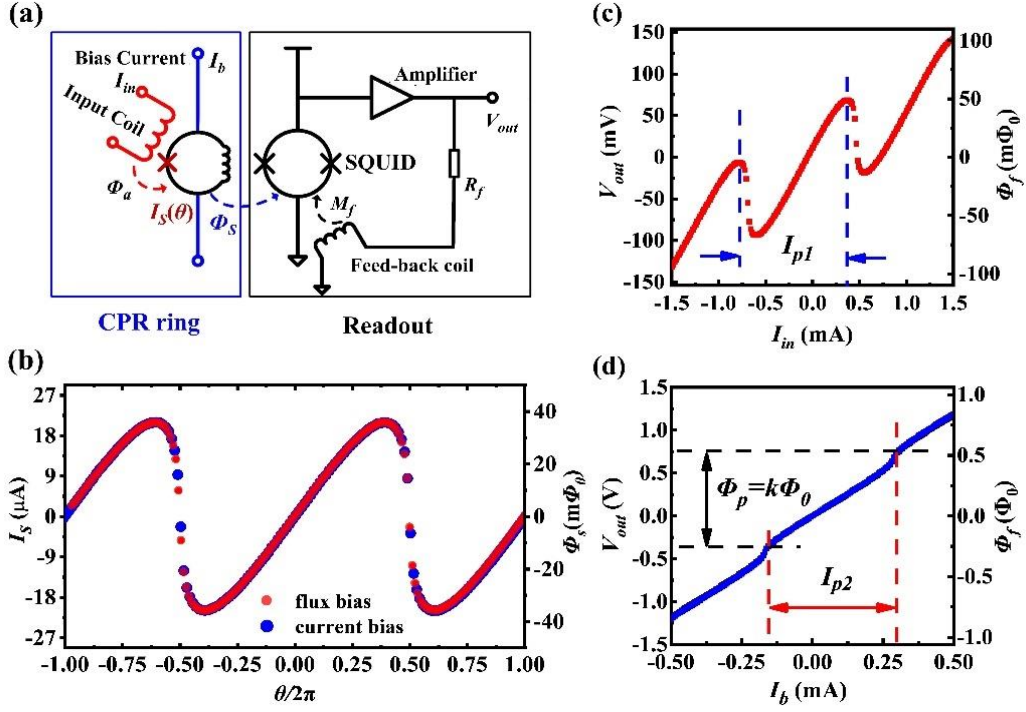

Figure 6. (a) Schematic of the CPR measurement. (b) CPRs of the nano-bridge derived from (c) and (d). (c) Typical measured curve by the flux-bias method. (c) Typical measured curve by the current-bias method.

Calculation of the CPR for NBJs using Ginzburg–Landau equations

The Ginzburg-Landau (GL) differential equations[37] are

$$\alpha\Psi + \beta|\Psi|^2\Psi + \frac{1}{2m^*}\left(\frac{\hbar}{i}\nabla - \frac{e^*\vec{A}}{c}\right)^2\Psi = 0, \quad (2)$$

$$J = \frac{e^*\hbar}{2m^*i}(\Psi^*\nabla\Psi - \Psi\nabla\Psi^*) - \frac{e^{*2}}{m^*c}\Psi^*\Psi\vec{A}. \quad (3)$$

where $m^* = 2m$ is the mass of the Cooper pair, $e^* = 2e$ is the charge of the Cooper pair, $\Psi$ is a complex order parameter describing the superconducting state, $A$ is the vector potential, and $J$ is the supercurrent density.

To solve the GL equations, we set $A = 0$ and present the complex function in the form $\Psi = |\Psi|e^{i\theta}$, where $|\Psi|$ is the modulus and $\theta$ is the phase. We set $f = |\Psi|/|\Psi|_0$, where $f$ is a function of the position coordinates $x$, $y$, and $z$, and $|\Psi|_0$ is the equilibrium value in a zero field. $\alpha = -\frac{\hbar^2}{2m^*\xi^2(T)} = -\beta|\Psi|_0^2$.

The real and imaginary parts of Eq. 2 are

$$\frac{1}{\xi^2(T)}f - \frac{1}{\xi^2(T)}f^3 + \frac{d^2f}{dx^2} - f\left(\frac{d\theta}{dx}\right)^2 + \frac{d^2f}{dy^2} - f\left(\frac{d\theta}{dy}\right)^2 + \frac{d^2f}{dz^2} - f\left(\frac{d\theta}{dz}\right)^2 = 0 \quad (4)$$

and

$$2\frac{d\theta}{dx}\frac{df}{dx} + f\frac{d^2\theta}{dx^2} + 2\frac{d\theta}{dy}\frac{df}{dy} + f\frac{d^2\theta}{dy^2} + 2\frac{d\theta}{dz}\frac{df}{dz} + f\frac{d^2\theta}{dz^2} = 0. \quad (5)$$

The supercurrent density $J$ is

$$J = \frac{e^*\hbar}{m^*}|\Psi|_0^2 f^2(\hat{x}\frac{d\theta}{dx} + \hat{y}\frac{d\theta}{dy} + \hat{z}\frac{d\theta}{dz}). \quad (6)$$

The differential equations Eq. 4 and 5 can be solved numerically by using of a finite element software.[38] The boundary conditions $\nabla\theta \cdot \hat{n} = 0$ and $\nabla f \cdot \hat{n} = 0$ are set on all boundaries except the two current entrance faces, where $\hat{n}$ is a vector normal to the surface. The phase is fixed at $-\theta_s/2$ at the left bank entrance face and $\theta_s/2$ at the right bank entrance face. $\theta_s$ is the total phase drop across the NBJ.

An example of the solutions for the phase $\theta$ and wave-function amplitude $f$ obtained using this model are shown in Figure 7. The nano-bridge is set to have a length $l = 30$ nm, width $w = 40$ nm, and thickness $d = 8$ nm. The phase drop across the nano-bridge is set at $\theta_s = 2.5$ rad. The distributions of $\theta$ and $f$ are shown in Figure 7 a and b, respectively. It is clearly seen that the phase gradient is localized over the region of bridge deck, and the bank serves as a good phase reservoir. The $X$, $Y$, and $Z$ components of current are calculated as $J_x = \frac{e^*\hbar}{m^*}|\Psi|_0^2 f^2 \frac{d\theta}{dx}$, $J_y = \frac{e^*\hbar}{m^*}|\Psi|_0^2 f^2 \frac{d\theta}{dy}$, and $J_z = \frac{e^*\hbar}{m^*}|\Psi|_0^2 f^2 \frac{d\theta}{dz}$. The total supercurrent $I_s$ passing through the NBJ is obtained by integrating $J$ over the current entrance face of the right bank:

$$I_s = \iint_s J_x\, dydz + J_y dxdz + J_z dydx = \iint_s J_x\, dydz. \quad (7)$$

By sweeping the value of $\theta_s$, we get the current-phase relations for different geometric dimensions, as shown in Figure 2b.

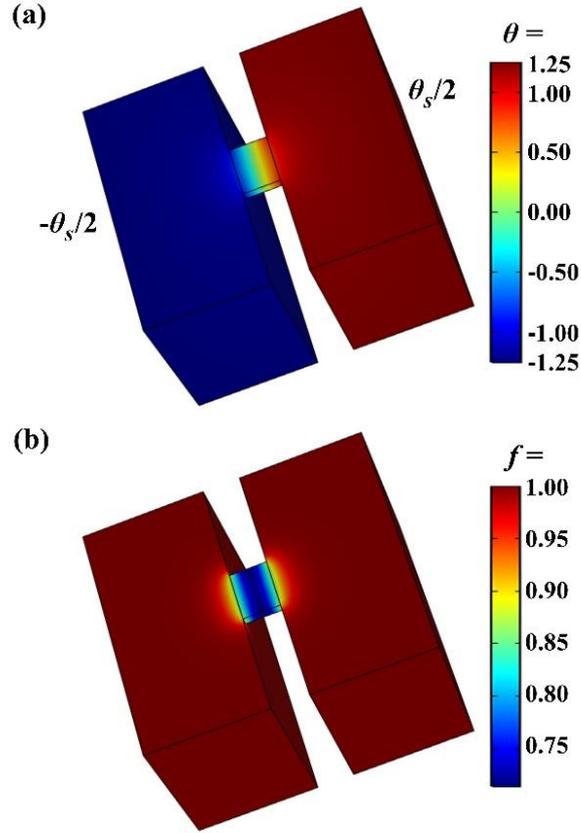

Figure 7. Solution of Ginzburg–Landau equations in the 3D bank structure: (a) wave function phase distribution for $\theta_s = 2.5$ rad and (b) wave function amplitude $f$ for $\theta_s = 2.5$ rad.

**Author Contributions**

Yue Wang and Lei Chen performed the experiments and collected the data. Lei Chen and Zhen Wang planned the research and wrote the paper. Yinping Pan, Denghui Zhang, Shujie Yu, and Guofeng Zhang participated and assisted the CPR measurements. Guangting Wu assisted in the CPR simulation work. Xiaoyu Liu, Ling Wu, Weifeng Shi, and Wei Peng assisted in the fabrication of devices. Lu Zhang, and Jie Ren assisted data analysis. All authors approved the final version of the manuscripts.

**Acknowledgements**

The authors acknowledge support from the National Natural Science Foundation of China (Grant No. 62071458, 11827805), the Strategic Priority Research program of CAS (Grant No. XDA18000000), and the Young Investigator program of the CAS (Grant No. 2016217). The fabrication was performed in the Superconducting Electronics Facility (SELF) of SIMIT.

## Additional Information

**Competing Interests:** The authors declare no competing interests

# SUPPORTING INFORMATION

1. **SEM images of devices for the CPR measurement**

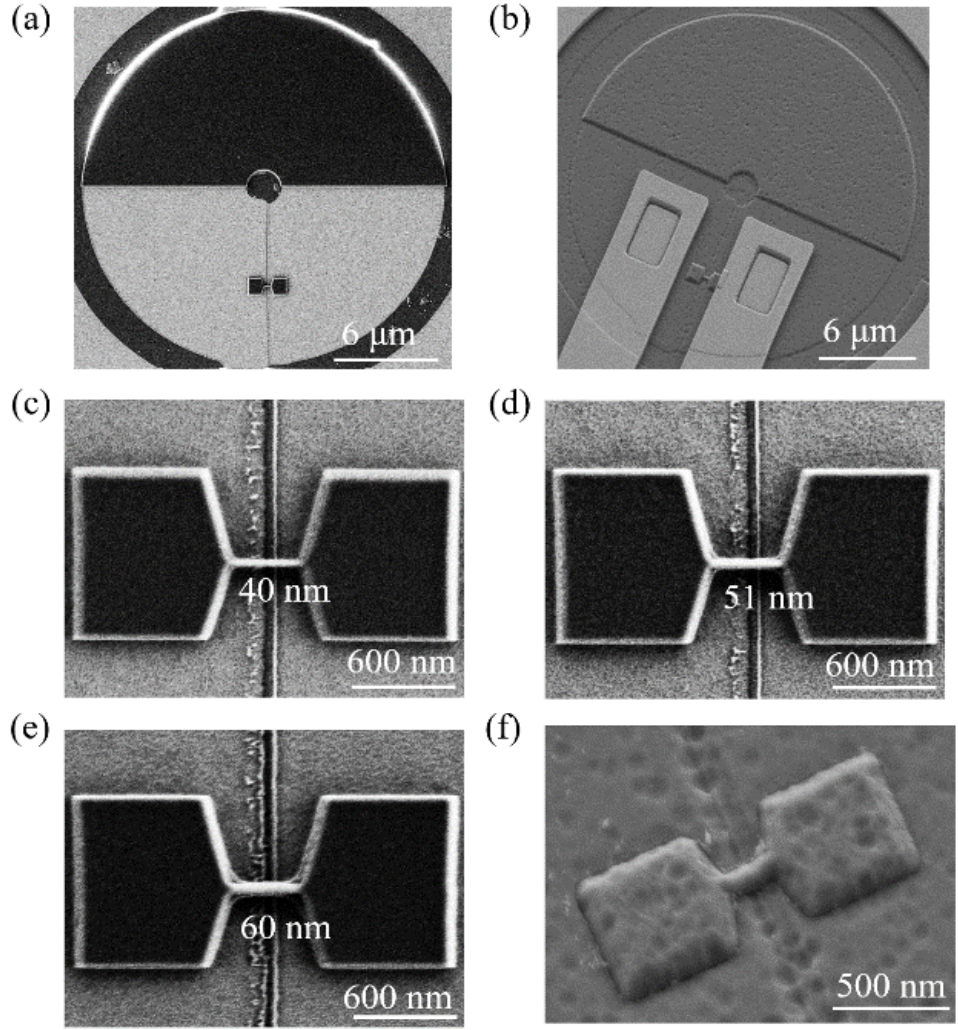

Figure S1 Scanning electron micrograph of the devices for the CPR measurement. The superconducting CPR ring with an NBJ (a) before covered by a $SiO_2$ insulating layer and (b) after the fabrication process was finished. (c), (d), and (e) the zoomed-in SEM images of the NBJs with the width of 40 nm, 51 nm, and 60 nm, respectively, before the deposition of $SiO_2$ layer. (f) The SEM images of the NBJ after the fabrication process were finished.

2. **TEM images of the cross-section of NBJs**

Figure S2 shows the TEM pictures of the cross-section of the insulating slit with the nano-bridge junctions (NBJs) on top, corresponding to the devices in Figure S1 (c-e). The thickness of the bridge is 5.2 nm, 6.6 nm, and 8.1 nm, respectively.

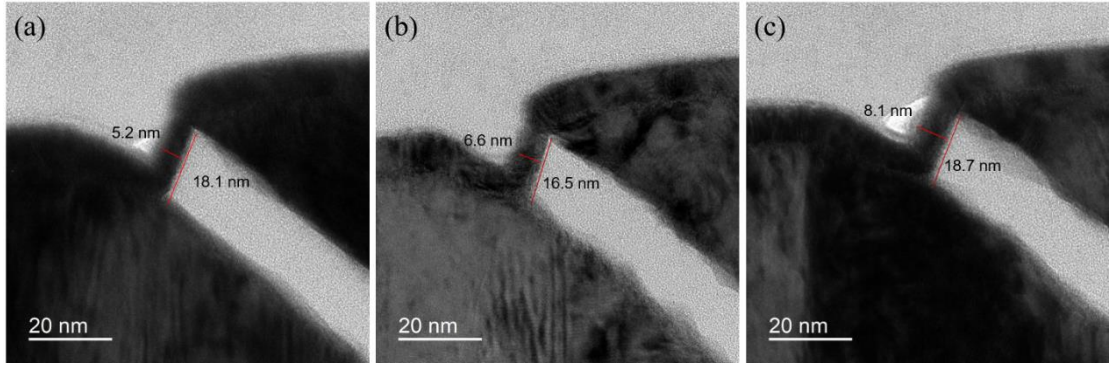

Figure S2 TEM pictures of the cross-section of the NBJs.

## 3. Multi-valued CPRs

The measured multi-valued CPRs are shown in Figure S3(a), which are corresponding to 4 points shown in the bottom right of Figure 2a for $\Delta\theta > 1$. The value of the $\Delta\theta$ is 1.57 rad (sample 1, red), 1.18 rad (sample 2, blue), 1.09 rad (sample 3, pink), and 1.20 rad (sample 4, green), respectively. Figure S3b plots several calculated CPR curves of NBJ with different sizes. The value of the $\Delta\theta$ is 1.55 rad (red, $l$ = 50 nm, $w$ = 40 nm, $d$ = 15 nm), 1.02 rad (blue, $l$ = 30 nm, $w$ = 50 nm, $d$ = 12 nm), 1.09 rad (pink, $l$ = 30 nm, $w$ = 60 nm, $d$ = 12 nm), and 1.21 rad (green, $l$ = 30 nm, $w$ = 60 nm, $d$ = 15 nm).

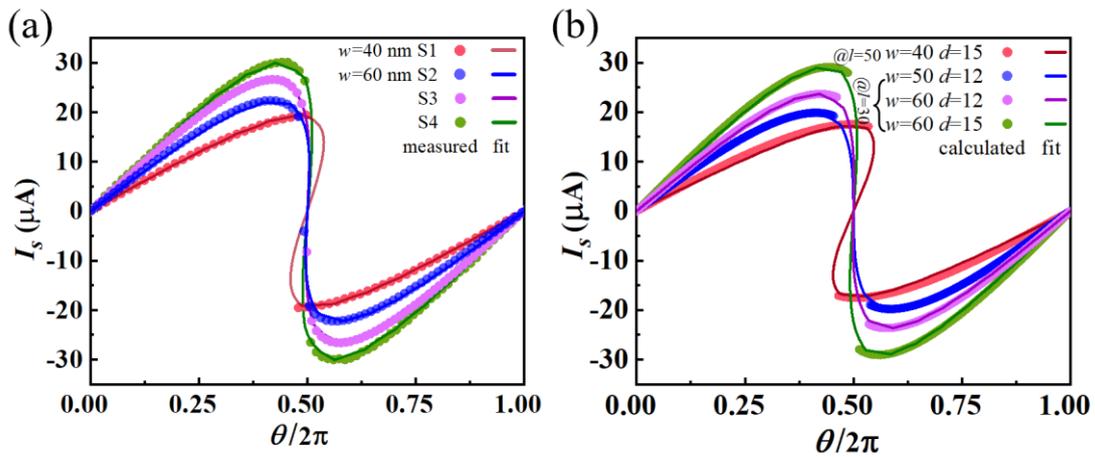

Figure S3 The measured (a) and calculated (b) multi-valued CPR curves. The units of $l$, $w$, and $d$ are nanometers